\documentclass[aps,prl,twocolumn,nofootinbib,preprintnumbers]{revtex4}
\usepackage{graphicx}  
\usepackage{dcolumn}   
\usepackage{bm}        
\usepackage{amssymb}   
\usepackage{comment}
\usepackage{amsmath,mathtools}
\usepackage{bbm}

\newcommand\ket[1]{| #1\rangle}

\usepackage[colorlinks=true, linkcolor=blue, urlcolor=blue, filecolor=black, citecolor=red, pdfstartview=FitV, bookmarksopen=true]{hyperref}

\usepackage{rotating}

\begin{document}

\title{Maximally entangling states and dynamics in one dimensional nearest neighbor Floquet systems }

\author{David Berenstein}
\email{dberens@physics.ucsb.edu}
\affiliation{Department of Physics, University of California, Santa Barbara, CA 93106, USA}

\author{Daniel Teixeira}
\email{dteixeira@usp.br}
\affiliation{Department of Physics, University of California, Santa Barbara, CA 93106, USA}
\affiliation{Institute of Physics, University of S\~ao Paulo, 05314-970 S\~ao Paulo, Brazil.}

\begin{abstract}

We describe conditions for generating entanglement between two regions at the optimal rate in a class of one-dimensional quantum circuits with Floquet dynamics. The optimal value follows from subadditivity and Araki-Lieb inequalities. A quantum circuit composed of parallel SWAP gates that act periodically on entangled pairs is a simple system that saturates the bound. We show that any other system that entangles at this maximal rate must act as a generalized SWAP gate dynamics on the relevant states of the Hilbert space. We further discuss some characterizations of states according to entropy generation. States with multipartite entanglement generically fail to entangle efficiently as time evolves. This suggests that chaos, which tend to produce such entanglement patterns, is expected to work against the process of spreading information efficiently. It also provides a simple intuition for why the entangling tsunami velocity must be slower than the Lieb-Robinson velocity.
\end{abstract}

\maketitle

The dynamical evolution of isolated quantum systems is unitary. This means that all the information encoded in their initial state can in principle be recovered by a clever experimenter. Such remark, however, does not apply to a subsystem of a larger system. Subsystems evolve and seem to thermalize, thereby effectively losing information  to the rest of the system as time passes by. Experimental progress recently achieved in the fine control of the environmental setup, in order to isolate quantum systems, has enabled implementing unitary time evolution to a good approximation for long time-scales. The possibility of addressing  long term open questions, such as the understanding of the dynamics leading to thermalization in experimental setups, has stimulated a wave of theoretical developments on its side. For recent reviews on both theoretical and experimental advances and perspectives about this vast subject, see \cite{Polkovnikov:2010yn, Langen:2014saa,Nandkishore:2014kca,Gogolin:2016hwy} and also references contained therein. 

A central role in the description of nonequilibrium phenomena is played by entanglement dynamics. This is particularly interesting in quantum field theories either on a lattice or in the continuum limit. The simplest nonequilibrium protocol in quantum field theory is a global quantum quench in $1+1$ dimensions. In this scenario, quasiparticle excitations propagate in an effective light-cone defined from the Lieb-Robinson bound \cite{Lieb:1972}, which also renders an upper bound on the generation of entanglement --- it can grow at most linearly with time \cite{Calabrese:2005in,Eisert:2006,Bravyi:2006}. In particular, the work of \cite{Calabrese:2005in} describes an intuitive model for entanglement entropy production that captures these results: the excitations are sourced from nearby points by the quench and after that they follow a ballistic journey with opposite quasimomenta. Freely propagating quasiparticles that are entangled display the linear behavior on entanglement entropy growth.

The aforementioned works deal with spin chains or two dimensional conformal field theories and they are valid for specific protocols. On the other hand,  for holographic field theory setups  that admit a classical gravity description, one can make considerable progress in computing the entropy dynamics via the holographic entanglement entropy proposals \cite{Ryu:2006bv,Hubeny:2007xt}. In general quantum field theories and protocols, however, the generation of entanglement does not have such a comprehensible overall view. Still, the lack of a complete picture compels us to study simplified models that  resemble quantum field theories. These models can be used to understand at least the phenomenology of  entanglement dynamics, for instance by providing a classification of quantum states according to entropy production. 

We will study a simple model of a lattice quantum field theory, given by certain constructions of quantum circuits with Floquet time evolution. By Floquet dynamics we mean that the system has a discrete unitary evolution which is periodic in time.
We highlight one example among this set, implemented through SWAP gates, that reproduces the picture of Cardy and Calabrese \cite{Calabrese:2005in}. This will provide some insights into different protocols for inhomogeneous quenches. This is a natural problem and of particular relevance, since initial states which are not translational invariant, as in local quenches, have concrete realizations in some experimental setups. For example, cold atoms subjected to a harmonic confinement or quantum wires subjected to a voltage difference \cite{Calabrese:2016xau}. More importantly, the setup considered here can be in principle efficiently simulated in a quantum computer.

\begin{figure}[ht!]
\begin{center}
		\includegraphics[scale=0.2]{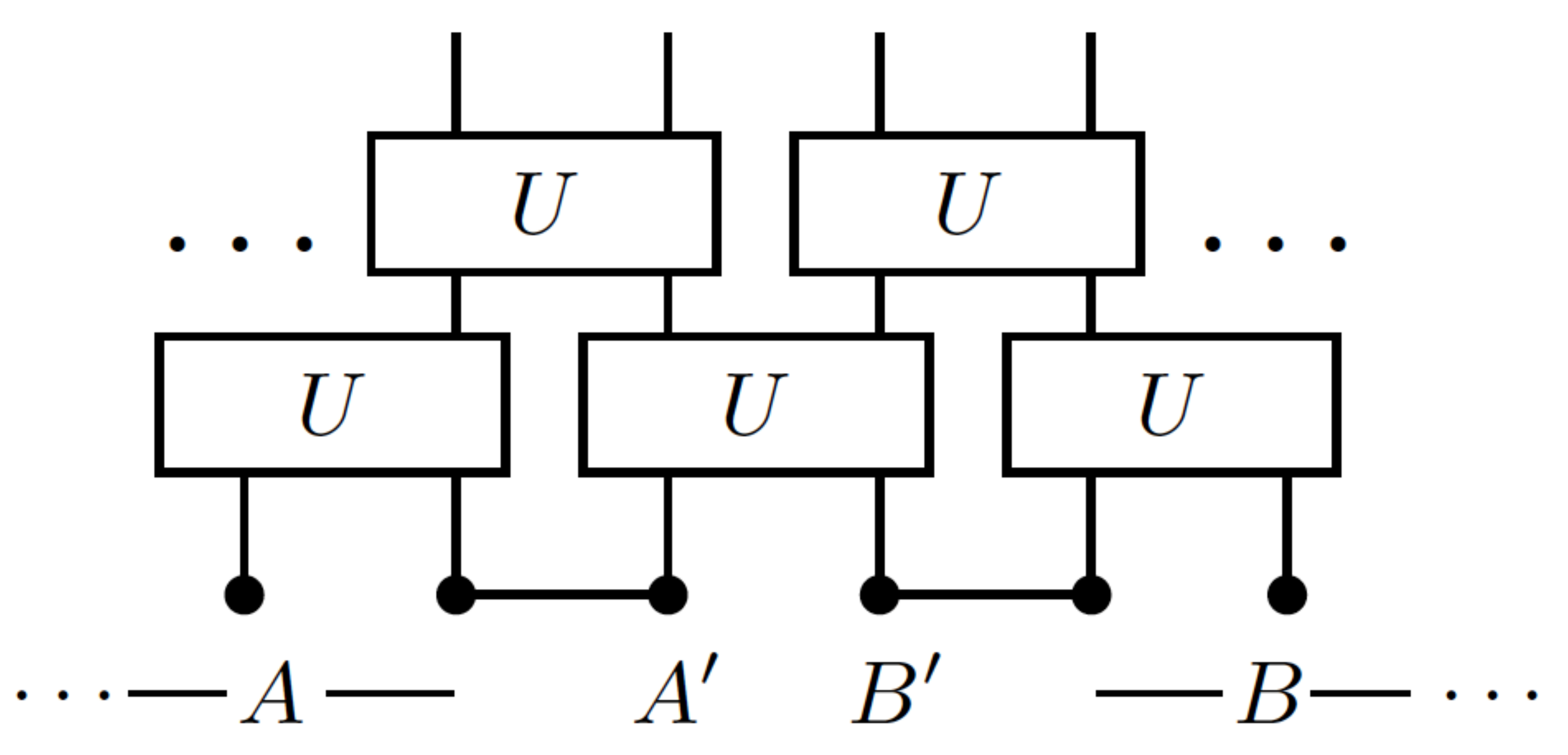} 
\caption{ \label{fig:circuit} Illustration of the quantum circuit under consideration. Dots represent lattice sites and segments joining two of them account for the fact that degrees of freedom on these sites are entangled.} 
\end{center}
\end{figure}

Let us start with a one-dimensional chain of $N$ sites, each with a Hilbert space $H_i$ of dimension $\dim(H_i)= M$ and implement the dynamics through the action of quantum gates $U$ at discrete times of length $\Delta t = 1/2$ (two layers are one unit of time). Such systems naturally realize a finite speed of propagation as would be required by the Lieb-Robinson bound. We shall consider two layers of unitaries connecting two sites, which operate on alternating nearest neighbors, and that are repeated after a period $T = 1$ producing a checkerboard pattern, as shown in Fig. \ref{fig:circuit}. The input state is pure and for simplicity it is prepared with entangled pairs between alternating adjacent sites, which is also displayed in Fig. \ref{fig:circuit} together with the partition we will consider in what follows. 

Such a system could be typically thought of as a quantum simulation of a quantum field theory in a system of qubits (or some other local Hilbert space), where $U$ is a discretization of the dynamical evolution obtained by some Suzuki-Trotter steps. 

We will be in general interested in protocols where two parts of the system, which we label as $AA'$ and $BB'$, begin in a pure state. Subsequently, they exchange information and become entangled as time evolves. The sites at which these two subsystems communicate are $A',B'$, while we take $A,B$ to be the rest of the system. Notationally, we have implicitly defined joined regions by adding labels, so that for example $AA' = A\cup A'$.
 
Subadditivity of the entanglement entropy together with the Araki-Lieb (or triangle) inequality read
\begin{equation}
\vert S_A(t)-S_{A'}(t) \vert \leq S_{AA'}(t) \leq  S_A(t)+S_{A'}(t),
\label{eq:ineq}
\end{equation}
for all times $t$, where we can assume $S_A(t)>S_{A'}(t)$ without loss of generality and remove the absolute value. In the rest of the discussion, we will always consider $t$ to be an integer. By construction, the unitaries acting at half-integer times do not mix regions $A$ and $A'$, such that $S_A(t) = S_{A}(t+1/2)$. Using this fact and (\ref{eq:ineq}), a bound on the entropy production at region $AA'$ during a period follows immendiately,
\begin{eqnarray}
\Delta S_{AA'} (t+1/2) &\equiv& S_{AA'}(t+1/2)-S_{AA'}(t) \notag \\ 
& \leq &  S_{A'}(t)+S_{A'}(t+1/2). \label{eq:bound}
\end{eqnarray}
The entanglement entropy bound is controlled by (a trivial constant times) the average entropy at site $A'$. Of course $\Delta S_{AA'}(t) = 0$, since the unitaries act within regions $AA'$ and $BB'$ separately. By symmetry, the same reasoning as above and the one that follows below applies once $AA'$ is replaced by $BB'$.

This bound is independent of the set of gates one has chosen to construct the circuit with, and it is saturated whenever $S_{AA'}(t+1/2)$ and $S_{AA'}(t)$ obey, respectively, subadditivity and Araki-Lieb inequalities with an equal sign. In the former case, the mutual information between $A$ and $A'$, $I_{A,A'}(t+1/2)$, is obviously zero meaning that $A'$ is completely entangled with $BB'$, while in the later event $I_{A',BB'}(t) = 0$ --- which holds since the input state is pure. It follows that $S_{AA'}(t) = S_{BB'}(t)$ and $S_{A'BB'}(t) = S_A(t)$ --- showing that $A'$ shares correlations only with $A$. In particular, strong subadditivity $I_{A',BB'}(t) \geq I_{A',B'}(t)$ implies that 
\begin{equation}
 S_{A'B'}(t) = S_{A'}(t) + S_{B'}(t).
 \label{eq:SAB}
\end{equation}
As a consequence, mutual information saturates a monogamy inequality $I_{A',BB'}(t) \geq I_{A',B}(t) + I_{A',B'}(t)$ in a trivial way. Necessary and sufficient conditions for the saturating Araki-Lieb inequality are classified in \cite{Carlen:2012}, for instance, by means of other measures of entanglement --- entanglement of formation and squashed entanglement.

Saturating both inequalities in (\ref{eq:ineq}) for many consecutive steps, and not just at one instant of time, is a nontrivial property of a system that depends both on the dynamics and on the initial state. When it holds, this leads to maximal entropy production at each step. Such feature motivate us to name a state as {\it maximally entangling} if it saturates both inequalities simultaneously for both partitions, $AA'$ and $BB'$, in the sense that it achieves the highest rate of entanglement entropy production that is allowed. A state can be maximally entangling at one instant of time or within an interval $[t_1,t_2]$. The idea is to understand what types of gates admit a large collection of maximally entangling states and how generic they are. We will call the full dynamical system {\em maximally entangling} if it admits a large class of generic states that are maximally entangling for long times. Here we provide an example of such a system through the choice of SWAP gates, whose action on any two states is given by SWAP$(\ket a\otimes \ket b) = \ket b\otimes \ket a$ when acting on two sites.

Thus, we now focus on the SWAP gate, where \\

\begin{figure}[ht!]
\vspace{-0.3cm}
\begin{center}
		\includegraphics[scale=0.13]{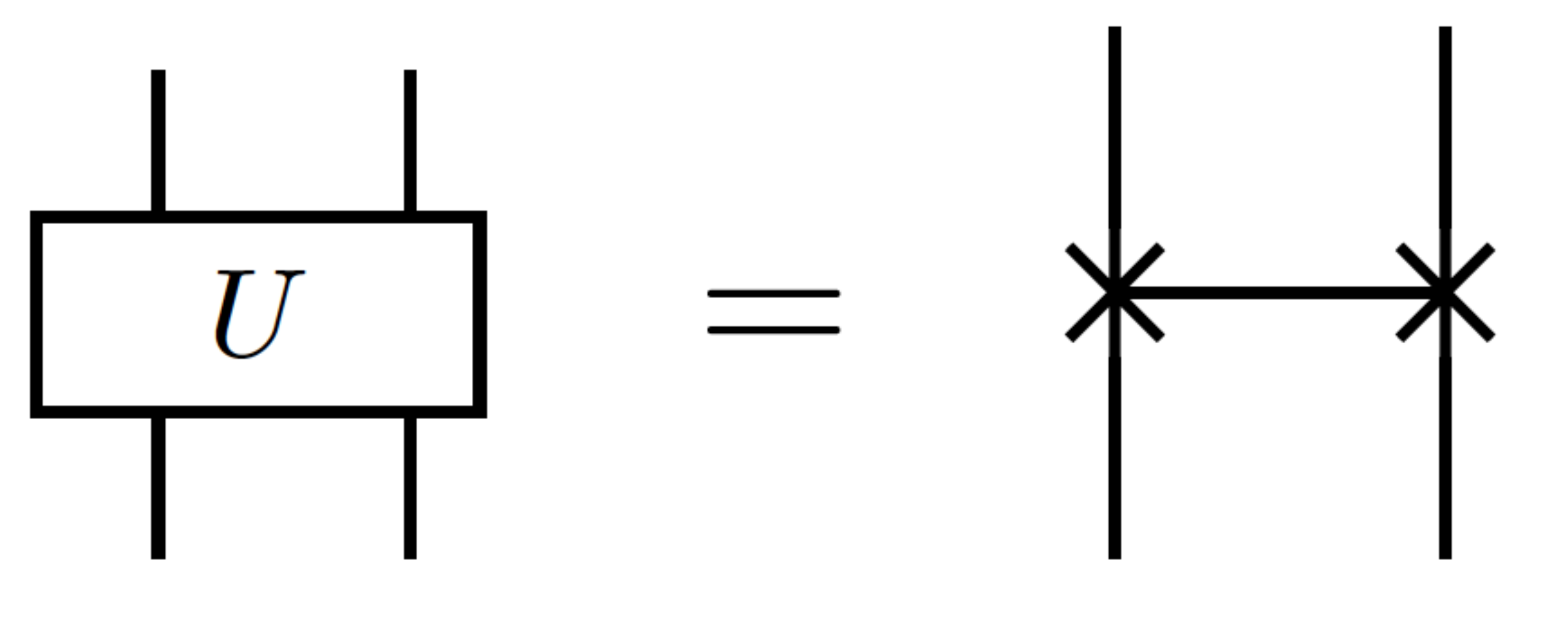}
\end{center}
\vspace{-0.4cm}
\end{figure}

\noindent The corresponding circuit is a model for a discrete non-interacting quantum field theory, where swapping two sites corresponds to a free streaming of particles to the left and right. This is an integrable model that can realize the intuition developed by Cardy and Calabrese in \cite{Calabrese:2005in}, where a quantum quench acts as a source of quasiparticle excitations that propagate freely with entangled left and right movers. Moreover, one can easily deal with input states that display an inhomogeneous entanglement pattern in this circuit model, which generalizes the free streaming picture to include a broader class of states.

In what follows, we will always take $N$ to be large and ignore finite size effects. Roughly speaking, we are taking $A,B$ to be infinite on a  first order approximation. 

Denoting the entanglement entropy at site $i$ at time $t=0$ by $s_i$ and assuming only nearest neighbor entanglement between pairs in the initial state (site $2k$ entangled only with site $2k+1$, but not necessarily maximally), it is straightforward to write down corresponding formulas for all regions. For instance, setting $A'$ to be at site $j$, one has
\begin{equation}
S_{A'}(t) = s_{j-2t}, \qquad S_{A'}(t+1/2) =s_{j+2+2t},
\end{equation}
where $t$ is an integer. In the above, we have also used the fact that for this class of states $s_{j+2+2t}=s_{j+1+2t}$. 

Also,
\begin{equation}
S_{A}(t) = s_j + \sum_{k=1}^t (s_{j-2k}+s_{j+2k})\label{eq:sat}
\end{equation}
and
\begin{equation}
S_{AA'}(t+1/2) = \sum_{k=0}^{t} (s_{j-2k} + s_{j+2+2k}).
\end{equation}
These equations clearly lead to the saturation of (\ref{eq:ineq}) for the relevant times discussed previously. Likewise, similar conclusions can be drawn for the $BB'$ side. In particular, since the SWAP model mimics the Cardy-Calabrese quench dynamics, this suggests that the Cardy-Calabrese setup should also saturate similar quantum information inequalities for entanglement generation. The quench should be maximally entangling in a certain sense for continuous time, rather than discrete time as above. This is a nontrivial result about entanglement production in a quantum field theory that can be inferred from the simple discrete approximation we are considering.

One can easily check further properties of the SWAP circuit. For instance, the entanglement between $A'$ and the rest of the system is swapped with the entanglement of $B'$ and the rest of the system at each step, in the sense that $S_{A'}(t+1/2) = S_{B'}(t)$ and, similarly, $S_{B'}(t+1/2) = S_{A'}(t)$. Moreover, for this class of initial states with bipartite entanglement, the tripartite information remains zero for all times,
\begin{equation}
I_{3}({A, B, B')} \equiv I_{A,BB'} - I_{A,B} - I_{A,B'} = 0,
\end{equation} 
meaning that the mutual information is always monogamous and extensive. This is of course expected since there is no generation of multipartite entanglement: information is only carried through the chain by the left and right movers which are correlated and remains localized, though in different locations. 

One can generalize the system to start in a pure state in $AA'$ and $BB'$ keeping the single site entropies fixed, $s_i=s_{A,B}$, differing only if they are in $A$ or $B$. One then expects based on Page's observation \cite{Page:1993df} that generically on each half $AA'$ and $BB'$, the left movers should be (maximally) entangled only with the right mover Hilbert space. This should be true even if $s_i$ is slowly varying on $AA'$ and $BB'$. Such generic states would evolve in a way that saturates the inequalities as well.

\begin{figure}[ht!]
\begin{center}
    \includegraphics[scale=0.18]{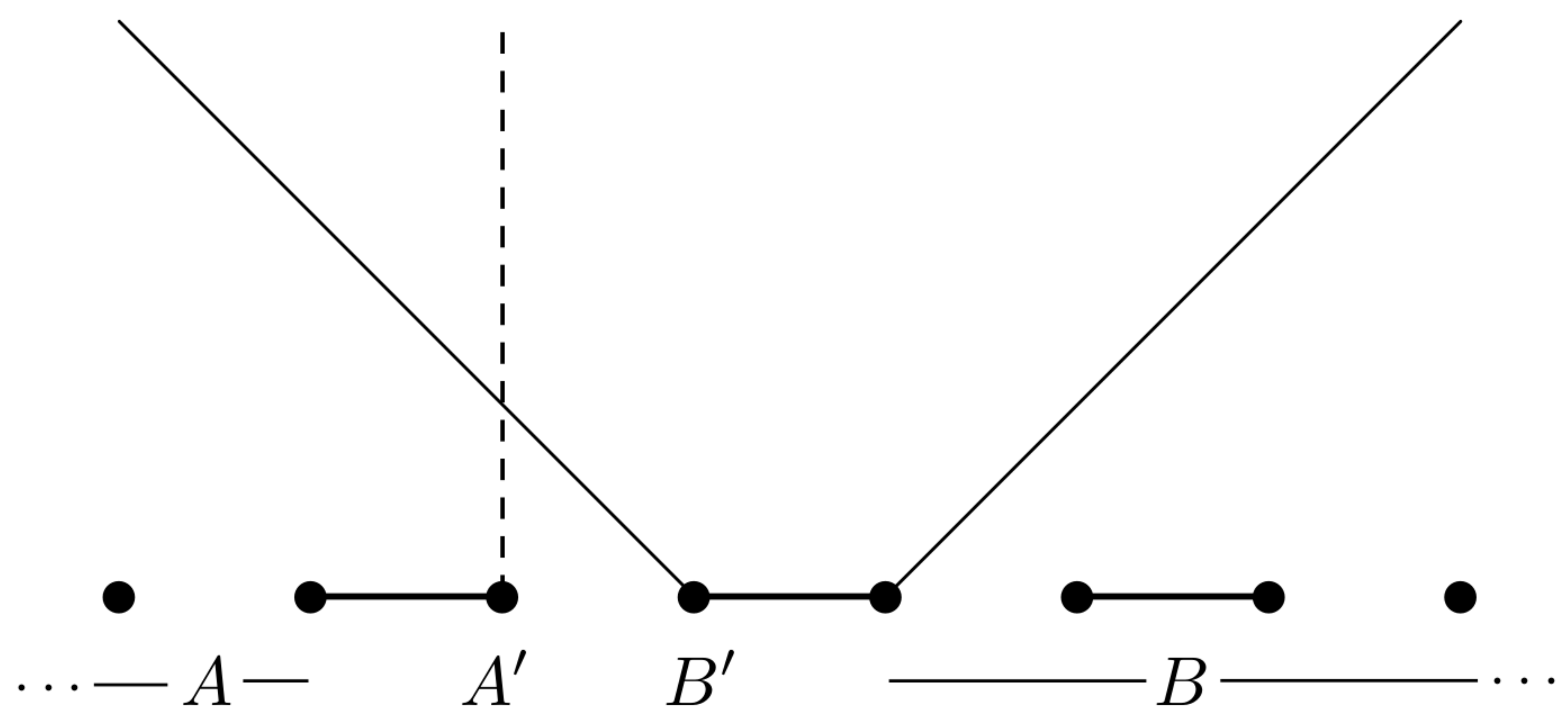}
    
    \vspace{0.5cm}
    
    \includegraphics[scale=0.19]{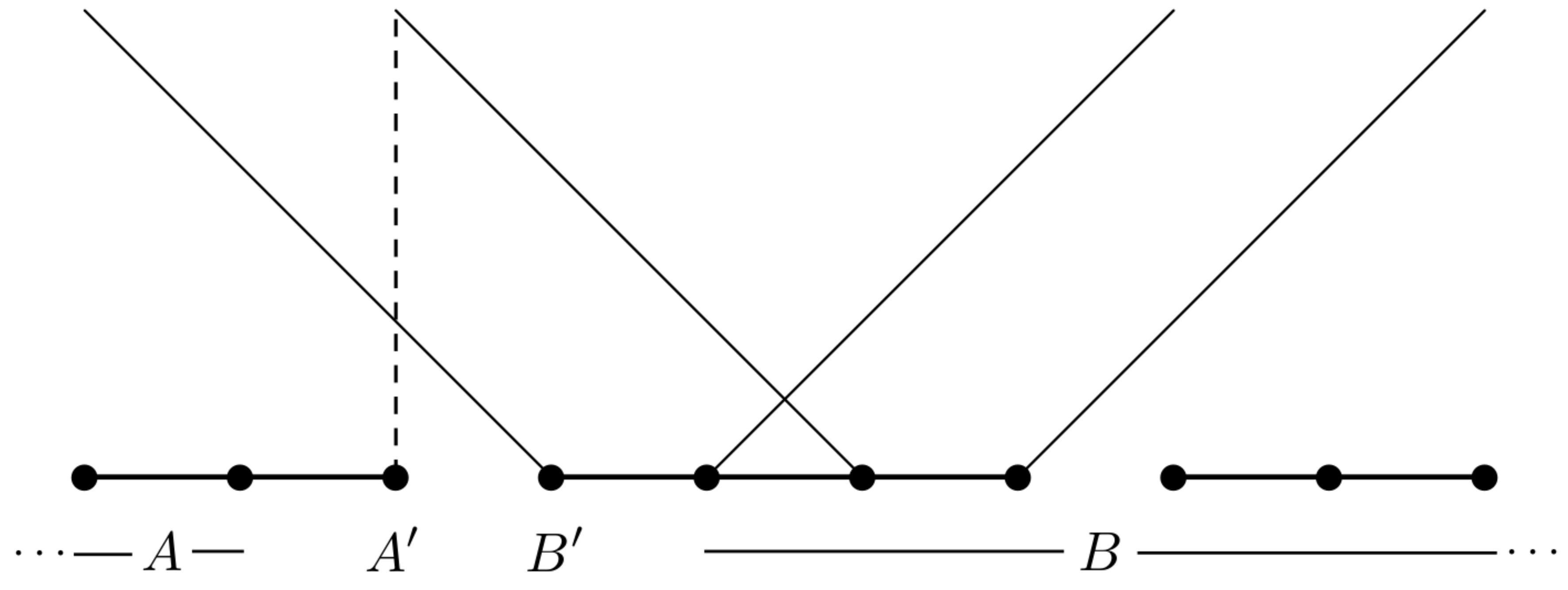}
\end{center}
\caption{ Streamlines followed by quasiparticles with bipartite entanglement (top) versus the multipartite case (bottom). In the former scenario, once a left-moving particle gets into $AA'$, it will give its maximum contribution to the entanglement entropy of this region. Alternatively, for a GHZ-like block of size $2k$, while there is an increase of $S_{AA'}$ once the leftmost particle enters $AA'$,  there is a delay proportional to $k$ before  additional entanglement between the two regions is generated due to the next block.}
\label{fig:ghz}
\end{figure}

On the other hand, states with multipartite entanglement, like GHZ states will fail to saturate (\ref{eq:ineq}) for many consecutive instants of time. It is easier to analyze this case for a system of qubits starting in a state of the form $a \ket{0}^{\otimes 2k}+b\ket{1}^{\otimes 2k}$, for definiteness. Then, even if there is initially an entanglement growth when the first particle crosses the interface between two subregions, the next set of particles will not contribute to increase the entropy further on, therefore such class of states will fail to saturate the bound. This happens because left movers are partially entangled with left movers in each region, $AA'$ and $BB'$, and not only with right movers as in the case with bipartite entanglement, as illustrated in Fig. \ref{fig:ghz}. 
Thus, states with multipartite entanglement will end up generating net entanglement between $AA'$ and $BB'$ at a slower rate. In other words, multipartite states are not maximally entangling.

This can be made very precise. Assume that there is some entanglement between left movers and left movers on side $BB'$ with some characteristic length $\xi$. Let us denote the postion at $B'$ as $i_1$. What this means for us is that if we have $\xi$ consecutive left-movers in $B$, then due to subaditivity of the entropy we have that $S_{i_1\dots i_\xi} \leq S_{i_1}+\dots + S_{i_\xi}$, but we insist on a strict inequality at distance $\xi$. As a consequence there is some mutual information between the left movers characterized by the distance scale $\xi$. One can then show that once the $\xi$-th qubit has traveled through the interface between $AA'$ and $BB'$, then equation \eqref{eq:sat} will not hold any longer, because $S_{AA'} (t+\xi) \leq S_{i_1\dots i_\xi}+S_{\rm rest} <s_j + \sum_{k=1}^t (s_{j-2k}+s_{j+2k})$.


The SWAP model provides insight into the role played by interactions in the process of generating entanglement. It shows that an integrable field theory already accomplishes the task of producing entanglement entropy at the optimal rate. By including chaos, if it has any effect on entangling regions at all, it must be the case that it slows the rate of entanglement growth: because entanglement production is saturated in the integrable case, it cannot increase further. Chaotic dynamics is notable for producing multipartite entanglement. Therefore, we conclude that chaos should mostly work against the process of spreading information efficiently. { We emphasize, however, that our reasoning does not rule out chaotic models to be maximally entangling. These models can still be maximally entangling as suggested for instance by the results obtained in \cite{Bertini:2018xow}. In fact, such example matches our expectation of realizing maximal rates for very specific dynamics --- self-dual points in the parameter space were considered in \cite{Bertini:2018xow} ---   at a very particular choice of initial states. At the same time, such system admits a quasiparticle description, like the integrable models, where our discussion will apply more generally (see the considerations concerning generalized SWAP gates below). }

It is worthwhile to notice that generic states that are maximally entangling in the SWAP model will fail to be so for different sets of gates. Cellular automata models with CNOT and ${\rm SL}(2,\mathbb{Z})$ gates were considered respectively in \cite{Berenstein:2018zif} and \cite{Zakirov:2018}, 
for instance, in the study of integrable and chaotic properties of many-body systems. The system of gates generically produce multipartite entanglement, such that our input state will not be maximally entangling according to those dynamics. Even more dramatically, if we use the CNOT dynamics, and we start with all states set to $\ket 0$ on the $AA'$ side, irrespective of what state we put on $BB'$, no entanglement between $A,B$ will be generated ever, even if the local entropy at $B'$ is maximal. This is because the system will evolve to states that are of the same form: a product of $\ket 0$ in $AA'$ times another state in $BB'$. For comparison, in general such states on $BB'$ would be maximally entangling by our criteria in the SWAP gate model.

Let us get back to the general set of gates and consider the case when there is no local entropy production beyond some time $t_{\ast}$.  This is one of the properties of the SWAP gate model. That is, we require that
\begin{equation}
S_{A'}(t_{\ast}) + S_{B'}(t_{\ast}) = S_{A'}(t_{\ast}+1/2) + S_{B'}(t_{\ast}+1/2).
\label{eq:loc-eq}
\end{equation}
{Such situation could arise, for instance, when the system reaches local equilibrium. Note that, by construction, mutual information between $A$ and $A'B'$ is conserved when we act only on $A'B'$, that is, $I_{A,A'B'}(t) = I_{A,A'B'}(t+1/2)$, and we can easily compute it for a maximally entangling system. Using $S_{AA'}(t) = S_A(t)-S_{A'}(t)$ together with (\ref{eq:SAB}) we get
\begin{eqnarray}
 \!\!\!\!\!\!\!I_{A,A'B'}(t)\!\! &=& \!S_A(t) + S_{A'B'}(t) - S_{AA'B'}(t) \notag \\
 &=& \! S_A(t) + S_{A'}(t) + S_{B'}(t) - S_{B}(t) \notag \\
 &=& \!S_{AA'}(t) + 2S_{A'}(t) - S_{BB'}(t) = 2S_{A'}(t). 
 \label{eq:2SA}
\end{eqnarray}
Here the local equilibrium \eqref{eq:loc-eq} does not play a role in this case. We will use it for half-integer times as well as $S_{AA'}(t+1/2)=S_A(t+1/2)+S_{A'}(t+1/2)$. Then, the corresponding steps to the ones above lead to
\begin{eqnarray}
 I_{A,A'B'}(t+1/2) = 2S_{B'}(t+1/2).
 \label{eq:2SB}
\end{eqnarray}
Hence, due to equality between \eqref{eq:2SA} and \eqref{eq:2SB}, and symmetry between $AA'$ and $BB'$, we have
\begin{equation}
  S_{A'}(t) = S_{B'}(t+1/2), \quad S_{B'}(t) = S_{A'}(t+1/2),
\end{equation}
meaning that, locally, entanglement in $A'B'$ is just swapped between the two sides at each time step when equilibrium holds. This suggests that, in these situations, $U$ is essentially a SWAP gate in the relevant states. The conditions \eqref{eq:loc-eq} must be achieved through a generalized SWAP gate, where we promote SWAP $\to\widetilde{\rm SWAP}$ = SWAP$\cdot (U_1\otimes U_2)$, for some one-site gates $U_1,U_2$, since in this case $U_1,U_2$ will not change the local entropy at each site. Then, ${\rm SWAP}\cdot\widetilde{\rm SWAP}$ acts as a product of unitaries on bipartite spaces. This statement is not true for the CNOT gate, for instance.}

In the context of the SWAP model, entropy conservation happens to be true even though energy is not conserved. However the system is integrable and has a lot of conserved quantities. In fact, \eqref{eq:loc-eq} holds for all periods. The physics responsible for this equilibration condition is a conservation law, which is usually energy conservation. If we consider the SWAP circuit with qubits, the conservation law can be taken to be the number of spins up, leading to a conserved spin density. Lack of entropy production can thus be thought of in terms of a conserved entropy current. Alternatively, this allows one to fix the maximal entropy density per unit of conserved charge as a proxy for temperature.

Now, returning to the field theory limit in $1+1$ dimensions, we expect a continuum version of \eqref{eq:bound} to take place, that is, 
\begin{equation}
\frac{{\rm d} S_A}{{\rm d} t}   \leq  \alpha {\mathbb \sigma (t)} ,
\label{eq:ent-prod}
\end{equation}
where ${\sigma (t)}$ is an entropy density and $\alpha$ has the units of velocity (or velocity times area in general dimensions). One should expect that this velocity is related to the velocity of propagation of interactions, defined from the Lieb-Robinson bound, which we will call $v_{LR}$. When the equality holds, $\alpha$ is  the so-called entanglement tsunami velocity \cite{Liu:2013iza}, $v_{E}$. This requires the assumption that the system thermalizes at late times, such that $\sigma(t)$ converges to some thermal entropy density. Since $v_{LR}$ is defined independent of an equilibrium hypothesis, we expect that $v_{E} \leq v_{LR}$, a result that can be proved for translation invariant states using causality arguments \cite{Casini:2015zua} or by considering the relative entropy with respect to a thermal state \cite{Hartman:2015apr}.

Recall that in the discrete case $\sigma(t)$ is proportional to the average entropy at a site. To get a finite entropy density in the limit where the lattice spacing goes to zero, we should consider situations where the local entropy per site is small. More precisely, since usually the entanglement entropy of a quantum field theory on an interval is divergent, we need $\sigma(t)$ to be a regularized entropy. For long enough times, we expect a system to achieve some notion of local equilibration, which sets a thermal scale given by the inverse temperature. Note that, while demanding low entropy per site naively suggests a slow entropy growth according to (\ref{eq:ent-prod}), it turns out that product states (where the state is pure at each site) can be instantaneously maximally entangling, no matter how small is the entropy production, as long as it is not zero. The idea is that since we start with zero entropy at each site, when we act with a non-trivial unitary on the boundary of the regions we are studying, one automatically saturates the bounds \eqref{eq:bound} because there is no mutual information between $A,A'$ and the state in $A$ is pure. Similarly for $B',B$. What this means is that the Lieb-Robinson velocity does not play a direct role in the instantaneous single gate model. To have such a bound, one needs enough time evolution so that the dynamical features of long distance  physics can be applied locally. This requires some notion of local equilibration. This argument shows that such an assumption, or a similar one, cannot be avoided in general.

As a special case, in a conformal field theory, the quantity $\sigma(t)$ can be related to the central charge $c$ in a (locally) thermal state by $\sigma \propto c T$, where $T$ is the temperature. The natural value for $\alpha$ is the Lieb-Robinson velocity: the speed of light in the conformal field theory. This suggests strongly that the Cardy-Calabrese quench is maximally entangling, since it saturates (\ref{eq:ent-prod}) with the conformal field theory values. While in $d=2$ we have $v_E = v_{LR} = 1$, this illustrates again that, in general, the entanglement tsunami velocity should then be smaller than or equal to the speed of propagation of fluctuations. The expectation from our analysis is that chaos is generically reducing the speed of entanglement between regions by increasing the multipartite entanglement among the microscopic degrees of freedom. This is in agreement with the results of \cite{Casini:2015zua}.

\medskip

\begin{acknowledgments}
\noindent {\em Acknowledgments ---} D. B. would like to thank extensive conversations with Xi Dong.  Work of D. B. supported by the Department of Energy grant DE-SC0019139. D. T. is grateful to the hospitality of UCSB during the development of this work. The research of D. T. was supported by FAPESP grant 2018/09330-7. 
\end{acknowledgments}

\end{document}